\documentclass[prd,aps,preprint,floats,nofootinbib,tightenlines]{revtex4}

\usepackage{epsfig}

\begin{document}
 
\pagestyle{empty}

\preprint{
\noindent
\begin{minipage}[t]{3in}
\begin{flushright}
\end{flushright}
\end{minipage}
}

\title{Inelastic final-state interaction}

\author{Mahiko Suzuki}
\affiliation{
Department of Physics and Lawrence Berkeley National Laboratory\\
University of California, Berkeley, California 94720
}

\date{\today}

\begin{abstract}

The final-state interaction in multichannel decay processes is systematically 
studied in the hadronic picture 
with application to $B$ decay in mind. Since the final-state interaction
is intrinsically interwoven with the decay interaction in this case, no simple 
phase theorem like ``Watson's theorem'' holds for experimentally observed final 
states. We first solve exactly the two-channel problem as a toy model in order 
to clarify the issues. The constraints of the two-channel approximation turns 
out to be too stringent for most $B$ decay modes, but realistic multichannel 
problems are too complex for useful quantitative analysis at present. To 
alleviate the stringent constraints of the two-body problem and to cope with 
complexity beyond it, we introduce a method of approximation that is 
applicable to the case where one prominent inelastic channel dominates over 
all others. We illustrate this approximation method with the amplitude of the 
decay $B\to K\pi$ fed by the intermediate states of a charmed-meson pair. 
Even with our approximation we need more accurate information of strong 
interactions than we have now. Nonetheless we are able to obtain some insight 
in the issue and draw useful conclusions on general features on the strong phases. 

\end{abstract}


\maketitle

\pagestyle{plain}

\setcounter{footnote}{0}

\section{Introduction}
    The well-known phase theorem \cite{Watson} holds for the 
final-state interaction (FSI) of decay processes when the final 
state consists of a single eigenstate of scattering. While no 
simple nontrivial extension is known in the case of multichannel final 
states, some calculations were made in the past with unjustified 
extension of the single-channel phase theorem\cite{early}. A two-channel
problem was studied with a certain class of $S$-matrix and the correct 
observation was made that inelastic channels are 
the main source of strong phases in many $B$ decay modes\cite{Don}. 
However, it is not easy to obtain quantitatively reliable results from the 
two-channel model. Taking the large limit of open channels, the statistical 
model \cite{CS,SW} was proposed as an alternative approach. 
Quantitatively, however, it is short of predictive power since it does 
not ask for detailed knowledge of strong interaction. When one approaches 
the problem in the quark-gluon picture, one faces inability or large
uncertainty in computing contributions of the soft collinear constituents 
numerically.  It is fair to say that at present we are far from successful 
computation of a FSI phase in multichannel decay processes. 

In this paper we first study the two-channel problem in detail.
The problem is solvable in a reasonably compact form without approximation
or assumption if relevant information is available about strong interaction 
physics and decay branching fractions.
The general solution to the two-channel toy-model shows how the elastic 
scattering phases and the channel coupling contribute to the total FSI phase. 
It concludes in agreement with Donoghue {\em et al}\cite{Don} that if a large
strong phase emerges in the $B$ decay into two light-mesons, its major 
source is coupling to decay channels that have large
branching fractions.\footnote{
This was suggested to the author by Wolfenstein in many occasions over 
years.\cite{W}} 
Although it points to the source of problems in strong phases, the simple 
two-channel model is inapplicable to $B$ decay.
We proceed to the case of more than two decay channels.  Even the 
three-channel problem is mathematically too complicated for solving 
in a compact form. On the physics side our knowledge of strong interaction 
at total energy 5 GeV ($\simeq m_B$) is not good enough to carry through 
the analysis with precision. To cope with formidable complexity of 
the problem, we introduce an approximation method that works in the case 
that, aside from the channel of our interest, one inelastic channel 
dominates over all others.  This is different from an approximate 
two-channel problem. 
It can happen, for instance, to the two-body light-hadron decays of $B$ 
meson when they couple to the charmed-meson pair states. We apply our 
approximation method to the decay $B\to K\pi$ and make semiquantitative 
analysis with knowledge of hadron physics currently available to us.   

\section{Framework and input}
    Two basic ingredients in discussion of FSI are unitarity and 
time-reversal. In the Standard Model the decay interaction is sum 
of effective local operators ${\cal O}_a (a=1,2,3,\cdots)$, each of which 
has a CP-violating (and therefore T-violating) phase factored out as
\begin{equation} 
 H_{int}e^{i\delta_{\rm w}}+ {\rm h.c.},
      \;\;\; (TH_{int}T^{-1}= H_{int}). \label{SM}
\end{equation} 
The T-violating ``weak phase'' $\delta_{\rm w}$ arises from the
CKM elements. In computing decay amplitudes we work separately on 
the $T$-invariant part $H_{int}$ of the decay interaction. 
In the case that $T$-violation is more general, we can break up the
weak interactions $H_w$ into $T$-even and $T$-odd parts as 
$H^{(\pm)}_{int}\equiv\frac{1}{2}(H_w\pm TH_wT^{-1})$ so that
\begin{equation}
       TH^{(\pm)}_{int}T^{-1}= \pm H^{(\pm)}_{int}.
\end{equation}
Then both $H^{(+)}_{int}$ and $iH^{(-)}_{int}$ are $T$-even since
$i\to i^*=-i$ under time reversal.\footnote{
If we make this breakup for the Standard Model interaction, we would 
get $H^{(+)}_{int}=\cos\delta_{\rm w}H_{int}$ and 
$iH^{(-)}_{int}=-\sin\delta_{\rm w}H_{int}$ for the first term 
$H_w = H_{int}e^{i\delta_{\rm w}}$ of Eq. (\ref{SM}).}
We should compute the decay amplitudes for 
$H^{(+)}_{int}$ and $iH^{(-)}_{int}$ separately and take a suitable
sum of them at the end.  Therefore it is sufficient to consider in general
only $T$-even weak interactions. No matter what method one may use, 
computation of FSI phases must always be made separately for different 
decay operators. Because two decay operators generate two different 
FSI phases for the same decay process even if the net quantum numbers of 
operators are identical. We consider in this paper only final-state 
interaction of strong interaction though it is in principle easy to 
include electromagnetic FSI. We shall refer to FSI phases also as strong 
phases in this paper.
  
With $T$-invariance, the strong $S$-matrix operator obeys
\begin{equation}
   TST^{-1} = S^{\dagger}.
\end{equation}
We can always choose phases of states such that the $T$-invariant 
$S$-matrix elements ($S_{kj}= \langle k|S|j\rangle = 
\langle k^{out}|j^{in}\rangle$) are not only unitary but symmetric;
\begin{equation}
     S_{jk} = S_{kj} \label{Symmetric}
\end{equation}
since $|j^{\rm in}\rangle\to\langle j^{\rm out}|$ and 
$\langle k^{\rm out}|\to|k^{\rm in}\rangle$ under time reversal. It 
is emphasized here that the requirement of Eq. (\ref{Symmetric}) fixes 
the phases of states except for the overall sign of $\pm 1$.\footnote{
If one multiplies the states with some 
phases as $|j\rangle \to e^{i\alpha}|j\rangle$ and $|k\rangle \to
e^{i\beta}|k\rangle$, $S_{jk}$ and $S_{kj}$ would acquire phases of
opposite signs $e^{\pm i(\beta-\alpha)}$ so that the equality 
$S_{jk}=S_{kj}$ would break down.}  
Specifically for the eigenchannels of the $S$-matrix $|a^{in}\rangle$ 
and $\langle b^{out}|$, it holds by definition that
\begin{equation}
  \langle b|S|a\rangle = \langle b^{out}|a^{in}\rangle
                       =\delta_{ba} e^{2i\delta_a}, \label{Eigen} 
\end{equation}  
where $\delta_a$ is the eigenphase shift. In the case of decay 
matrix elements, the initial state is a one-particle state that is
stable with respect to strong interaction. Since the initial decaying state 
is an asymptotic state with respect to strong interaction, there 
is no distinction between ``in'' and ``out'' states. For $B$ decay
\begin{equation}
        |B\rangle \stackrel{T}{\longrightarrow} \langle B| \label{B},
\end{equation}
where we choose the $B$ meson at rest. Eq. (\ref{B}) removes an arbitrary 
unphysical phase from the state $|B\rangle$ too.

A simple relation results from time reversal of the decay matrix 
element $\langle a^{out}|H_{int}|B\rangle$ when the final state
$\langle a^{\rm out}|$ is an eigenstate of $S$. T-invariance of
$H_{int}$ leads to
\begin{equation}
  \langle a^{out}|H_{int}|B\rangle
 =\langle B|H_{int}|a^{in}\rangle. \label{T}
\end{equation}
Inserting the completeness relation 
$\sum_b|b^{out}\rangle\langle b^{out}|=I$ next to $H_{int}$ in the 
right-hand side, we obtain from Eq. (\ref{T}) with Eq. (\ref{Eigen})
\begin{equation}
 \langle a^{out}|H_{int}|B\rangle
 =\sum_b\langle B|H_{int}|b^{out}\rangle\langle b^{out}|a^{in}\rangle
 = e^{2i\delta_a} \langle a^{out}|H_{int}|B\rangle^*.    
\end{equation} 
With the decay amplitude $\langle a^{out}|H_{int}|B\rangle$ into
the eigenchannel $a$ denoted by $A_a$, this relation reads  
\begin{equation}
        A_a = e^{2i\delta_a}A_a^*,     \label{Ph}
\end{equation}
namely, $A_a = \pm e^{i\delta_a}|A_a|$. 
This is the well-known phase theorem usually referred to as Watson's
theorem\cite{Watson}. It is a powerful theorem when the final state 
is an eigenstate of $S$. It has been an important tool of analysis
in hyperon decay and $K\to \pi\pi$ decay. 

But this relation is of little use when rescattering has 
inelasticity. We mean by {\em inelasticity} that observable final 
states are linear combinations of eigenstates whose weights are 
{\em not} simply determined by the Clebsch-Gordan coefficients
of isospin symmetry or SU(3) coefficients. In such cases the phase 
of the elastic scattering amplitude has little to nothing to do 
with the FSI phase of the corresponding decay amplitude, as we 
shall see it in a moment.

Usefulness of the phase theorem is thus limited to the decay of low-mass 
particles where rescattering is purely elastic up to isospin structure.
If the $K$ meson mass were sufficiently above 1 GeV, for instance, 
$\pi\pi$ of definite isospin would no longer be an eigenstate of 
$S$-matrix even approximately. The state $\rho\rho$ and $\omega\omega$
would enter an $S$-matrix eigenchannel with $\pi\pi$ and composition of 
such an eigenstate depends on low-energy dynamics of the transition
among $\pi\pi$, $\rho\rho$, and $\omega\omega$. In the case of $B$ 
decay an experimentally observed final state is a linear combination 
of many different $S$-matrix eigenstates so that the net FSI phase 
results from the eigenphases weighted with the decay amplitudes of 
$B\to {\rm eigenstate}$. Take. for instance, the $\pi\pi$ final state 
in $I=0$ of $\overline{B}^0$ decay. The state $|\pi\pi\rangle_{I=0}$ 
(in $s$-wave) is far from being an $S$-matrix eigenstate at energy 
$m_B$. If we want to use the phase theorem Eq. (\ref{Ph}), we must 
expand $|\pi\pi\rangle$ in the strong $S$-matrix eigenstates at $m_B$. 
However, we have little knowledge of these eigenstates since their 
composition depends sensitively on strong interaction at long and 
intermediate distances.  Experimentally the two-body channels do not 
account for all final states in $B$ decay. Three and four particle 
final states of the same $J^{PC}$ may be significant. Unless there is 
a good reason to believe that channel coupling is negligible among 
these final states, the strong $S$-matrix eigenstates at $m_B$ are 
made of many different particle states ($\pi\pi$, $\rho\rho$, 
$K\overline{K}$, $\pi\pi K\overline{K}$, $\cdots$). Therefore, 
if we expand the $\pi\pi$ state at total energy $m_B$, it is a linear 
combination of many different eigenstates of strong $S$-matrix. We 
would have to know these expansion coefficients in order to determine 
the FSI phase of $\pi\pi$ final state with the phase theorem.

Let us formulate what we have described above. When an observable 
final state $|i^{\rm out}\rangle$ is not $S$-matrix eigenstate, 
we expand it in the eigenstates of $S$-matrix $|a^{\rm out}\rangle$ 
as\footnote{
We represent the $S$-matrix eignestates by $|a\rangle$, $|b\rangle, 
\cdots$ and the observed particle states by $|i\rangle$, $|j\rangle$,
$|k\rangle$, $\cdots$.}
\begin{equation}
       |i^{\rm out}\rangle = \sum_a O_{ia}|a^{\rm out}\rangle .
                     \label{Expansion}
\end{equation} 
We choose that the $S$-matrix is symmetric ($S_{ij}=S_{ji}$) in the 
basis of the observable states. (cf Eq./ (\ref{Symmetric})) Then the 
expansion coefficients $O_{ia}$ are real, that is, the transformation 
matrix ${\cal O}$ in Eq. (\ref{Expansion}) is an orthogonal matrix. 
(See Appendix if proof is needed.) With this expansion and Eq. (\ref{Ph}),
\begin{equation}
     A_i = \sum_a O_{ia}A_a = \sum_a O_{ia}e^{2i\delta_a}A_a^*.
                              \label{Multichannel}
\end{equation}
We are able to write the right-hand side of Eq. (\ref{Multichannel}) 
in terms of observable decay amplitudes $A_j$'s as
\begin{eqnarray}
     A_i &=& \sum_{aj} O_{ia}e^{2i\delta_a}O_{ja}A_j^*, \nonumber\\
         &=& \sum_j S_{ij}A_j^*, \label{FSI}
\end{eqnarray}
where $S_{ij}= O_{ia}e^{2i\delta_a}O_{ja}$ has been used. This is the  
fundamental relation in discussion of FSI. The physical picture is simple 
and clear. (Fig.1) The input is unitarity and time-reversal of the 
$S$-matrix aside from choice of unphysical phases of states. Dynamical 
information of strong interaction is fed through 
the eigenphases $\delta_a$ and the orthogonal mixing matrix ${\cal O}$. 
In addition, we must provide relative magnitude of the amplitudes $A_i$ 
as independent pieces of input from weak interaction.  Consequently 
the phase of $A_i$ depends not only on strong interaction but also on 
weak interaction. It is clear here that the FSI phase of a decay amplitude 
has virtually nothing to do with the phase of the elastic scattering 
amplitude $a_J(s)$ of $J=0$ at 5 GeV. Although it looks almost futile 
to go any further, the purpose of this paper is to extract something 
useful for B decay out of Eq. (\ref{FSI}). 

\begin{figure}[h]
\epsfig{file=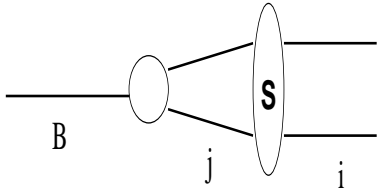,width=5cm,height=2.5cm} 
\caption{The final-state interaction relation in diagrams. Eq. (12).
\label{fig:1}}
\end{figure}
 
\section{Dynamical input}
 
We focus on the FSI phases of the two-body decay modes of the $B$ 
meson. The phases of three-body decay amplitudes depend on the 
sub-energies of three particles.  It is only the phases integrated 
over the sub-energies with the total energy fixed to $m_B$ that 
enter Eq. (\ref{FSI}).  We need the $S$-matrix elements $S_{ij}$
from experiment. To be concrete, let us consider the elastic 
$\pi^+\pi^-$ scattering amplitude as an example. The argument below 
is identical for other two-body channels of light mesons. In $B$ 
decay the relevant partial-wave channel is $\pi^+\pi^-$ in $s$-wave 
($J^P=0^+$) with isospin $I=0$ and $I=2$. 

Although the high-energy $\pi\pi$ scattering cannot be directly 
measured in experiment, we can make a reasonable estimate about 
elastic $\pi\pi$ scattering since the center-of-momentum energy 
5 GeV is in the high-energy asymptotic region well above the 
$\pi\pi$ resonances.\footnote{
The excited charmonia exist at mass not far below 5 GeV but their 
coupling to light hadrons is suppressed by QCD.  Coupling of 
$\pi\pi$ to the open charm channels\cite{Charming} is one of interesting 
subjects of our study later in this paper.}
The two-body light-hadron scattering in the high energy asymptotic 
region was studied theoretically and experimentally in the 1960's.
The Regge theory describes this physics well. The properties 
of the Regge trajectories can be deduced from meson-nucleon and 
nucleon-nucleon scattering even though we have no meson-meson 
scattering experiment. We give a very brief review\cite{Collins} 
of our knowledge in this area of the past time since it is an 
important input in our study of FSI.  

First of all, the Pomeron
exchange dominates in high-energy elastic scattering. (Fig 2) The
Pomeron may include a cut and can be more a complicated singularity
than a simple pole in $J$-plane unlike the non-Pomeron trajectories
such as $\rho$ and $f_2$. 
At the level of numerical accuracy of our discussion, however, we 
treat the Pomeron as a simple pole at $J=\alpha(t)$ with the intercept 
$\alpha(0)=1$ and a vanishingly small slope.  This entails factorization 
of the $J$-plane residue into product of two vertices. Since isospin is 
zero for the Pomeron, it contributes equally to the $I=0$ and $I=2$ 
states of the crossed channels $\pi\pi$. With these properties of the 
Pomeron we obtain necessary pieces of information on $\pi\pi$ scattering 
from $\pi p$ and $pp$ scattering at high energies.

\begin{figure}
\epsfig{file=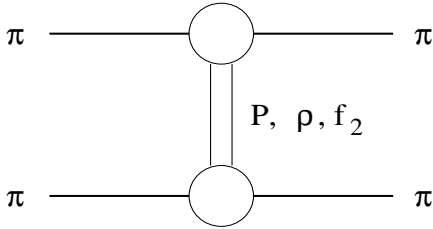,width=6cm,height=3cm} 
\caption{The Regge exchange in elastic $\pi\pi$ scattering. 
\label{fig:2}}
\end{figure}

The invariant amplitude for the asymptotic elastic scattering is
parametrized with the Regge parameters of the Pomeron in the form
\begin{equation}
        A^{\pi\pi}(s,t) = -\beta_P^{\pi\pi}(t)
 \frac{1+e^{-i\pi\alpha_P(t)}}{\sin\pi\alpha_P(t)}
       \Big(\frac{s}{s_0}\Big)^{\alpha_P(t)}
                                         \label{Inv}
\end{equation} 
where $\beta_P^{\pi\pi}(t)=(\beta_P^{\pi}(t))^2$ and $\alpha_P(t)\simeq 
1+\alpha'_P(0)t +\cdots$ are the factorized residue and the trajectory 
of the Pomeron, respectively, in terms of the invariant momentum transfer 
$t$.\footnote{
The possible small $\log^2s$ rise of $\sigma_{tot}(s)$ does not set 
in at $\sqrt{s}\simeq $5 GeV.  We may safely ignore it in our numerical 
work. We have included in $\beta_P^{\pi\pi}(t)$ the $t$-dependence of some 
other kinematical factors such as $2\alpha_P(t)+1$. The value of 
$s_0$ is at our choice, normally chosen to be 1 GeV$^2$.}
We can fix $\beta_P^{\pi\pi}(0)$ by the optical theorem $\sigma_{tot}
={\rm Im}A(s,0)/s$ and the factorization relation 
$\sigma_{tot}^{\pi \pi}\simeq (\sigma_{tot}^{\pi p})^2/\sigma_{tot}^{pp}
\simeq 22mb$ at $\sqrt{s}=m_B$ \cite{Collins,PDG}. The value $22mb$ 
is in line with the empirical quark counting rule, $\sigma_{tot}^{\pi\pi}
\simeq\frac{2}{3}\sigma^{\pi p}_{tot}\simeq (\frac{2}{3})^2
\sigma_{tot}^{pp}$\cite{Levin}. The width of the forward peak is more 
or less universal to all elastic hadron scatterings\cite{Collins2,Rarita}. 
We approximate the Pomeron slope to zero ($\alpha_P'(0)\simeq 0$) and 
fit the forward elastic peak with the standard exponential form
$\exp(t/t_0)$. The diffraction peak parameter is fixed by experiment 
to $t_0\simeq (0.22\sim 0.29){\rm GeV}^2$ by the elasticity
$\sigma_{el}/\sigma_{tot}$ at $\sqrt{s}=m_B$\cite{PDG,LBL0,LBL}.  
This value of $t_0$ 
reproduces the $t$ dependence of $d\sigma_{\rm el}/dt$ that falls by 
roughly three orders of magnitude from $t=0$ to $t\simeq -1\; {\rm GeV}^2$ 
for $pp$ and $\pi p$ scattering\cite{Collins2,Rarita}. It is 
justified by factorization to choose the $\pi\pi$ forward peak parameter 
equal to that of $pp$ and $\pi p$. The $\pi\pi$ invariant amplitude 
at high energies is thus set to
\begin{equation}
  A^{\pi\pi}(s,t)\simeq 22\;{\rm mb}\times i\;s\; e^{\frac{t}{t_0}},
\end{equation}
where we choose $t_0 = (0.253\pm 0.033){\rm GeV}^2$. The uncertainty in $t_0$ 
is primarily due to whether one estimates it with $\sigma_{el}/\sigma_{tot}$ 
of $\pi^{\pm}p$ or $pp$ and $p\overline{p}$. 

The partial-wave amplitudes $a_l(s)$ can be projected out
of $A^{\pi \pi}(s,t)$.  The result for the $s$-wave is:
\begin{equation}
     a_0^{\pi \pi}(s) = (0.282\pm 0.037) i, \;\; ({\rm at}\;s=m_B^2), 
                            \label{partial}
\end{equation}
which leads to the $s$-wave  $S$-matrix with 
$2|{\bf p}_{\rm cm}|/\sqrt{s}\simeq 1$, 
\begin{equation}
     S_0^{\pi\pi}(s) = 1 +2ia_0^{\pi\pi}(s).
\end{equation}
Hereafter we shall often parametrize strength of elastic scattering 
by $\epsilon$ as 
\begin{equation}
     S_0(s) = 1 - \epsilon, \label{epsilon}
\end{equation}
With Eq. (\ref{partial}), the Pomeron contribution to the $S$-matrix 
of $\pi\pi$ scattering at $m_B$ is
\begin{equation} 
     S_0^{\pi\pi}\simeq 1 - (0.56\pm 0.07). \label{pi} \label{Spi}
\end{equation}
The partial-wave amplitudes $a_l(s)$ extracted from the flat Pomeron 
amplitude are purely imaginary for all $l$. The amplitude $a_0(s)$
approaches asymptotically the imaginary axis below the center of the 
Argand diagram (shown schematically for $I=0$ in Fig. 3). 
If one described the high inelasticity of 
high-energy $\pi\pi$ scattering by an absorptive black sphere potential, 
one would have $S_0^{\pi\pi}(s)\to 0$ ({\em i.e.,} $a_0(s)\to 0.5i$). 
In this limit $\sigma_{\rm el}\to\frac{1}{2}\sigma_{\rm tot}$ for all 
$l$'s by shadow scattering effect, which is in disagreement with experiment. 
Although the numerical value in the right-hand side of Eq. (\ref{Spi}) 
has been extracted for the $\pi\pi$ channel, it is much the same for 
other two-meson channels. With a help of the $Kp$ cross section 
$\sigma_{tot}^{Kp}$\cite{PDG} we obtain 
\begin{eqnarray}
    &S_0^{\pi K} \simeq&  1- 0.51, \;\;(\pi K) \nonumber\\ 
    &S_0^{K\overline{K}} \simeq&  1- 0.45, \;\;(K\overline{K}),
                 \label{K}
\end{eqnarray}
where uncertainties are comparable to $\pm 0.07$ quoted for $\pi\pi$
in Eq. (\ref{Spi}) or a little larger.

\begin{figure}
\epsfig{file=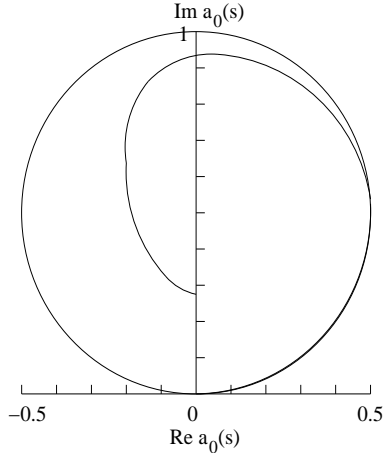,width=5cm,height=6cm} 
\caption{Energy dependence of $a_0(s)$ in the complex plane (Argand diagram). 
\label{fig:3}}
\end{figure}

The values in Eqs. (\ref{pi}) and (\ref{K}) are the Pomeron contribution 
alone. The nonleading Regge exchanges generate a small imaginary part for
$S_0$. The relevant trajectories are the $\rho$ and $f_2$ in the case of
$\pi\pi$. Their contributions can be estimated with a few additional 
theoretical inputs\footnote{
The inputs are the isospin-current coupling of $\rho$ and the exchange 
degeneracy of $\rho$ and $f_2$.}
from the cross section difference $\sigma_{\pi^-p}-\sigma_{\pi^+p}\simeq
1.6 mb$ at $\sqrt{s}=m_B$\cite{LBL}. Within the uncertainty due to  
$t$-dependence of the residue $\beta(t)$, their contributions to 
${\rm Im}S_0$ are at the level of $0.05i$ for $\pi\pi$.  Some may 
wonder about validity of extracting the $s$-wave amplitude from the 
forward peak region alone. The $s$-wave amplitude has a flat angular 
dependence so that the contribution of $a_0(s)$ extends equally to all 
directions ($P_0(\cos\theta)=1$). On the other hand experiment shows 
that the forward peak falls off by more than three orders of magnitude 
and there is no sign of the $s$-wave contribution at large angles. 
But this is no surprise. The $s$-wave amplitude at large angles is 
canceled by the partial-wave amplitudes of up to $l=O(\sqrt{s})$ 
which are rapidly oscillatory in angular dependence as 
$P_l(\cos\theta)\sim \sin[(l+\frac{1}{2})\theta+
\frac{\pi}{4})]/\sqrt{l\sin\theta}$ (for $l\theta\gg 1$). 

Now our task is to extract useful pieces of information from 
Eq. (\ref{FSI}) with the high-energy elastic $S$-matrix of Eq. 
(\ref{pi}) or Eq. (\ref{K}).         
  
\section{Final state interaction of two coupled channels}

  Let us first count how many dynamical quantities are involved in the 
most general $(n\times n)$ FSI relation, Eq. (\ref{FSI}).  The unitary 
and symmetric $S$ matrix contains $\frac{1}{2}n(n+1)$ independent parameters;
$n$ eigenphase shifts and $\frac{1}{2}n(n-1)$ rotation angles of
${\cal O}$. To solve for $A_i$'s of observable channels in Eq. (\ref{FSI}), 
therefore, we must feed the $\frac{1}{2}n(n+1)$ dynamical parameters of 
strong interaction. This is not sufficient to determine $A_i$ uniquely.
Although the FSI relation Eq. (\ref{FSI}) may look as if it introduced
$2n$ constraints through the real and imaginary parts, only a half of 
them, namely $n$ of them are actually independent.\footnote{
In the case of a single channel, the relation $A_1 = e^{2i\delta_1}A^*_1$ 
gives a constraint only on the phase of $A_1$, not its magnitude. In the 
case of $n$ channels, something similar happens: The phases of $A_a$ for 
{\em eigenchannels} are determined when $S_{ij}$ are completely specified, 
but their magnitudes $|A_a|$ are not.}  
We must provide the relative magnitudes of $A_a$ or $A_i$ as an additional 
input in order to determine the FSI phases uniquely. The magnitude of 
a decay amplitude is determined primarily by weak interaction, {\em i.e.,} 
the property of decay operators. Knowledge of strong interaction alone
can never determine multichannel FSI phases. We need to know interplay
of strong and weak interactions.

  Nobody is capable of tackling this problem for a general value $n$.
We will therefore be content with studying the FSI relation first in the
simple manageable case of $n=2$ and then searching a sensible approximation 
in more complicated and realistic cases. 
 
   Although the two-channel problem is the next to the simplest, there 
has been no serious attempt to study this case in the past, probably with
a good reason as we see below. Although it may not look much 
relevant to the $B$ decay of the real world, we have a chance to see
through general characteristics of coupled channel effects. For instance, 
how is the FSI phase of $\pi\pi$ channel affected by the $\rho\rho$ channel ? 
If one of the charm-anticharm channels such as 
$D^{(*)}\overline{D}^{(*)}$ strongly couples to the $\pi\pi$ channel, 
how does this channel affect the FSI of the $\pi\pi$ channel ?\footnote{
For some dynamical reason the branching fraction to $\rho\rho$ is 
an order of magnitude larger than that to $\pi\pi$. The $D^*\overline{D}^*$
channel has a huge branching fraction because of the robust 
$b\to c$ transition.} 
While simple-minded perturbative calculations have been undertaken
in the past, we would like to study these questions 
systematically with the two-channel toy model that incorporates unitarity. 

     We can write the general $T$-invariant $S$-matrix of $2\times 2$ 
with three parameter ($\frac{1}{2}n(n+1)=3$ for $n=2$) in the form of
\begin{eqnarray}
     S &=& \left( \begin{array}{cc}
            \cos\theta & -\sin\theta \\
            \sin\theta &  \cos\theta \end{array}\right)
         \left( \begin{array}{cc}
             e^{2i\delta_1} & 0   \\
             0 & e^{2i\delta_2} \end{array}\right)
         \left( \begin{array}{cc}
            \cos\theta & \sin\theta \\
           -\sin\theta & \cos\theta  \end{array}\right),         
                                           \nonumber \\
        &=& \left( \begin{array}{cc}
            e^{2i\delta_1}c_{\theta}^2 + e^{2i\delta_2}s_{\theta}^2 &
            (e^{2i\delta_1}-e^{2i\delta_2})c_{\theta}s_{\theta} \\
            (e^{2i\delta_1}-e^{2i\delta_2})c_{\theta}s_{\theta} &
            e^{2i\delta_1}s_{\theta}^2 + e^{2i\delta_2}c_{\theta}^2, 
                   \end{array} \right) \label{Two}
\end{eqnarray}  
where $\cos\theta$ and $\sin\theta$ are abbreviated as $c_{\theta}$ and 
$s_{\theta}$ in the second line. Substituting this $S$-matrix in the FSI 
relation Eq. (\ref{FSI}), we obtain the constraints on the real and 
imaginary parts or the magnitudes and phases of the decay
amplitudes defined by\footnote{
Recall that the arbitrary unphysical phases of states have been fixed up 
to an overall $\pm$ sign by the symmetry condition $S_{ij}=S_{ji}$ on 
the $S$-matrix.}
\begin{equation}
A_j = a_j + ib_j =|A_j|e^{\Delta_j}, \;\; (j=1,2) \label{Delta}
\end{equation} 
where the phases $\Delta_{1,2}$ are the FSI phases (the strong phases) 
of channel 1 and 2. We have in mind $j=1$ for $\pi\pi$ and $j=2$ 
for either $\rho\rho$ or $D^{(*)}\overline{D}^{(*)}$ of $I=0$.  
The constraining equation of Eq. (\ref{FSI}) can be written out for 
the real and imaginary parts as
\begin{equation}
      \left( \begin{array}{c} a_1 \\ a_2 \\ b_1 \\ b_2 \end{array} \right)
   = \left( \begin{array}{cc}
        {\rm Re}S & {\rm Im}S \\
        {\rm Im}S & - {\rm Re}S \end{array} \right)
     \left( \begin{array}{c} a_1 \\ a_2 \\ b_1 \\ b_2 \end{array} \right).
\end{equation}
where ${\rm Re}S$ and ${\rm Im}S$ are the $2\times 2$ matrices.
To be explicit,
\begin{equation}
  \left( \begin{array}{c}
       a_1 \\ a_2\\b_1\\b_2 \end{array} \right) =
  \left( \begin{array}{cccc}
           c_{2\delta_1}c_{\theta}^2+c_{2\delta_2}s_{\theta}^2&
           (c_{2\delta_1}-c_{2\delta_2})c_{\theta}s_{\theta}&
           s_{2\delta_1}c_{\theta}^2+s_{2\delta_2}s_{\theta}^2&
           (s_{2\delta_1}-s_{2\delta_2})c_{\theta}s_{\theta}, \\
                                        \nonumber
          (c_{2\delta_1}-c_{2\delta_2})c_{\theta}s_{\theta} &
           c_{2\delta_1}s_{\theta}^2+c_{2\delta_2}c_{\theta}^2 &
          (s_{2\delta_1}-s_{2\delta_2})c_{\theta}s_{\theta} &
           s_{2\delta_1}s_{\theta}^2+s_{2\delta_2}c_{\theta}^2 \\
                                           \nonumber
           s_{2\delta_1}c_{\theta}^2+s_{2\delta_2}s_{\theta}^2 &
          (s_{2\delta_1}-s_{2\delta_2})c_{\theta}s_{\theta} &
         -(c_{2\delta_1}c_{\theta}^2+c_{2\delta_2}s_{\theta}^2) &
         -(c_{2\delta_1}-c_{2\delta_2})c_{\theta}s_{\theta}, \\
                           \nonumber
          (s_{2\delta_1}-s_{2\delta_2})c_{\theta}s_{\theta} &
           s_{2\delta_1}s_{\theta}^2+s_{2\delta_2}c_{\theta}^2 &
         -(c_{2\delta_1}-c_{2\delta_2})c_{\theta}s_{\theta} &
         -(c_{2\delta_1}s_{\theta}^2+c_{2\delta_2}c_{\theta}^2)
                 \end{array} \right)
          \left( \begin{array}{c}
           a_1 \\ a_2\\b_1\\b_2 \end{array} \right).
           \label{Relation2}  
\end{equation}
As we have pointed out above, Eq. (\ref{Relation2}) contains only two 
independent constraints, not four. Indeed, one can show that two 
eigenvalues of the $4\times 4$ matrix in the right-hand side are unity 
and generate no constraint. The two remaining eigenvalues are $-1$ 
and generate constraints. 

     We now feed our dynamical input Eq. (\ref{epsilon}) in the
slightly different notation;
\begin{equation}
      S_{11}= 1-\epsilon,\;\;\; (|S_{11}|<1) \label{S11}
\end{equation}
where $\epsilon$ is real but not necessarily very small in magnitude.
For our discussion later we choose $\epsilon$ to be $\simeq 0.5$, 
as suggested by the Pomeron dominance in elastic scattering. It is easy 
to include the nonleading Regge contributions and relax the condition 
$\epsilon^* = \epsilon$. Fixing $\epsilon$ amounts to setting two parameters
in $S$-matrix of Eq. (\ref{Two}) so that we are left with one parameter 
out of three. When we fix the $S_{11}$ component as in Eq. (\ref{S11}), 
it is more convenient to parametrize the $S$-matrix in the form
\begin{equation}
    S  = \left(\begin{array}{cc}
 1-\epsilon & 
  i\sqrt{2\epsilon -\epsilon^2}e^{i\chi}\\
  i\sqrt{2\epsilon -\epsilon^2}e^{i\chi} & 
      (1-\epsilon)e^{2i\chi} \end{array} \right), \label{Two'}
\end{equation}
where the angle $\chi$ is the remaining single parameter of the 
$S$-matrix. It is related to two eigenphase shifts $\delta_{1,2}$  
defined in Eq. (\ref{Two}) by
\begin{equation}
   \sin 2\chi
  =\frac{\sin^2 2\delta_1-\sin^2 2\delta_2}{
     \sin2(\delta_1-\delta_2)}.                    
\end{equation}
Note that once $S_{11}$ is given, magnitude of the channel coupling 
$|S_{12}|=\sqrt{2\epsilon-\epsilon^2}$ in Eq. (\ref{Two'} is fixed 
by unitarity and no longer a free parameter.

\subsection{Case of $|S_{22}-1|>|S_{11}-1|$}

Let us first study the case where the partial-wave amplitude 
$a_0^{\pi\pi}(s)$ of rescattering is stronger in the second channel 
than in the first channel, namely, {\em i.e.,} $|S_{22}-1| > |S_{11}-1|$.  
This may serve as a toy-model of $\pi\pi$ and 
$D^*\overline{D}^*$;\footnote{
The $D^*\overline{D}^*$ branching is larger than that of $D\overline{D}$.
The $D^*\overline{D}$ and $D\overline{D}^*$ channels cannot make $J^P=0^+$.}  
The $D^*\overline{D}^*$ scattering of $I=0$ is presumably strong because of 
the near-threshold enhancement and/or broad excited charmonium resonances.  
For illustration we consider the extreme case that the rescattering 
in the second channel is maximally strong relative to that in the first
channel. This is realized by choosing $e^{2i\chi}= -1 (\chi = \frac{\pi}{2})$ 
in Eq. (\ref{Two'}). The symmetric unitary $S$-matrix takes the form of
\begin{equation}
   S = \left( \begin{array}{cc}
          1-\epsilon & -\sqrt{2\epsilon-\epsilon^2} \\
 -\sqrt{2\epsilon-\epsilon^2} & -1+\epsilon \end{array} \right),
           \;\; (\epsilon^* = \epsilon).
                        \label{S2} 
\end{equation} 
In terms of partial-wave amplitudes, this gives $a_0^{\pi\pi}(s)=
\frac{1}{2}i \epsilon$ and $i(1-\frac{1}{2}\epsilon)\approx \frac{3}{2}i
\epsilon$ for the first and 
second channels, respectively, with $\frac{1}{2}\epsilon\approx 0.25$. 
By substituting this $S$-matrix in the FSI relations, 
we obtain (see Eq. (\ref{Delta})) 
\begin{eqnarray}
  A_1 &=& a_1 + ib_1, \\ \nonumber
  A_2 &=& -\sqrt{\frac{\epsilon}{2-\epsilon}}\biggl(a_1 
                  -i\frac{2-\epsilon}{\epsilon}b_1 \biggr).
                              \label{A}
\end{eqnarray} 
The real and the imaginary parts of $A_1$ are still independent of each
other. The phases $\Delta_{1,2}$ cannot be determined uniquely even after 
the $S$-matrix is fully specified. But the phase $\Delta_1$ is related 
to $\Delta_2$ by
\begin{eqnarray}
   \tan\Delta_1 &=& - \frac{\epsilon}{2-\epsilon}\tan\Delta_2
                \nonumber \\
                &\simeq & - \frac{1}{3} \times \tan\Delta_2. \label{A'}
\end{eqnarray}   
Even when $\chi$ of $S_{22}=|S_{22}|e^{2i\chi}$ is equal to 
$\frac{1}{2}\pi$ (normally called as ``resonant''), $\Delta_2$ is not 
necessarily equal to $\frac{1}{2}\pi$. Although this may look puzzling 
at the first sight, it is not.  To determine $\Delta_{1,2}$ uniquely, 
we need to feed one more piece of information.  For instance, if the 
value of the ratio $|A_2/A_1|$ is supplied, we can determine $\Delta_1$ 
and $\Delta_2$ individually. By eliminating $\Delta_2$ from 
Eq. (\ref{A}) we obtain the relation that determines $\Delta_1$ in 
terms of $|A_2/A_1|$ and $\epsilon$:
\begin{equation}
   \sin^2\Delta_1 =\frac{\epsilon}{4(1-\epsilon)}
   \biggl[(2-\epsilon)\frac{|A_2|^2}{|A_1|^2} - \epsilon\biggr].  
     \label{Delta1}
\end{equation}
The ratio $|A_2/A_1|$ contains information of weak interactions. 
In multichannel decay, weak interaction plays a very important role 
in determining the FSI phases.
 
There is one shortcoming of this two-channel toy model:
As one sees in Eq. (\ref{Delta1}), the ratio $|A_2/A_1|$ must lie 
in the range of
\begin{equation}
  \sqrt{\frac{\epsilon}{2-\epsilon}}\leq \frac{|A_2|}{|A_1|}
      \leq \sqrt{\frac{2-\epsilon}{\epsilon}} \label{Range}
\end{equation}
for this model to be applicable.
At the lower boundary of $|A_2/A_1|$, it happens that $\Delta_1=
\Delta_2=0$, while $\Delta_1= -\Delta_2 = \pm 90^{\circ}$ at the
upper boundary of $|A_2/A_1|$.  When $|A_2/A_1|$ is in between, both
 $\Delta_1$ and $\Delta_2$ take nonzero values even though all elements
of the $S$-matrix are real.  This is an important point to be
emphasized. The phase of $A_1$ can arise from the process $B\to 2\to 1$ 
through the intermediate state $2$.  Some may wonder why the ratio 
$|A_2/A_1|$ is constrained in the two-channel toy model. 
With channel coupling present, one channel feeds the other by FSI to 
the direction to equalize magnitudes of $|A_1|$ and $|A_2|$. The FSI 
not only generates phases for $A_1$ and $A_2$ but also alters their 
magnitudes. Highly asymmetric $|A_1|$ and $|A_2|$ are incompatible 
with the FSI connecting the two channels unless $\epsilon\to 0$, 
{\em i.e.,} $|S_{12}|\to 0$.

\subsection{Case of $S_{22}\simeq S_{11}$} 

  When channel 2 is $\rho\rho$, we expect that the elastic $\rho\rho$ 
scattering is asymptotic at $\sqrt{s}\simeq 5$GeV and very similar to 
the elastic $\pi\pi$ scattering:
\begin{equation}
     S_{22} = S_{11} = 1-\epsilon, \;\;(\epsilon^* = \epsilon) 
\end{equation} 
In this case unitarity and symmetry require that the off-diagonal
element $S_{12}$ should be purely imaginary ($\chi=0$ 
in Eq. (\ref{Two'})): 
\begin{equation}
    S = \left( \begin{array}{cc}
          1-\epsilon & \sqrt{2\epsilon-\epsilon^2}i \\
 \sqrt{2\epsilon-\epsilon^2}i & 1-\epsilon \end{array} \right).
                                \label{S2'}
\end{equation}
This is the case that was studied by Donoghue {\em et al}\cite{Don}. 
The FSI relation Eq. (\ref{FSI}) leads us to
\begin{eqnarray}
    A_1 &=& a_1 +ib_1 \nonumber \\
    A_2 &=& \sqrt{\frac{2-\epsilon}{\epsilon}}\biggl(b_1
               +i\frac{\epsilon}{2-\epsilon}a_1\biggr),  \nonumber \\
  \tan\Delta_1\tan\Delta_2 &=&\frac{\epsilon}{2-\epsilon}
        \simeq \frac{1}{3}.    \label{Solution2}
\end{eqnarray}
Here again the FSI phases are uniquely determined only after the
ratio $|A_2|/|A_1|$ is given. The phase $\Delta_1$ is expressed in 
terms of $|A_2/A_1|$ and $\epsilon$ by the same relation as Eq. 
(\ref{Delta1}). The ratio $|A_2/A_1|$ lies in the same range 
as Eq. (\ref{Range}). In contrast to the case of the maximum 
$|S_{22}-1|$, the amplitude $A_2$ and $A_1$ are $90^{\circ}$
out of phase, one real and the other purely imaginary, at the 
lower and upper boundaries of the range for $|A_2/A_1|$.

\subsection{General two-channels}
     Once we have explored the two extreme cases above, it is not 
difficult to find the general solution for an arbitrary value of
$\chi$ in Eq. (\ref{Two'}). By rewriting the FSI relation in $A_2' 
\equiv e^{-i\chi}A_2$, we can reduce it to the case of $S_{22}=S_{11}$
above. We obtain the solution for general $\chi$ as
\begin{eqnarray}
    A_1 &=& a_1 +ib_1 \nonumber \\
    A_2 &=& \sqrt{\frac{2-\epsilon}{\epsilon}}e^{i\chi}\biggl(b_1
     +i\frac{\epsilon}{2-\epsilon}a_1\biggr),  \nonumber \\
    \tan\Delta_1\tan(\Delta_2-\chi) &=& \frac{\epsilon}{2-\epsilon}.
                        \label{General}  
\end{eqnarray}
The first and second lines of Eq. (\ref{General}) reduce to those of the 
previous cases; $|1-S_{22}|={\rm max}$ and $S_{11}=S_{22}$ as $\chi\to 
\frac{1}{2}\pi$ and $\chi\to 0$, respectively. The third line relates 
the FSI phases $\Delta_1$ and $\Delta_2$ to each other with parameter
$\epsilon$ and $\chi$. The ratio $|A_2/A_1|$ is related to $\Delta_1$ 
and $\epsilon$ exactly in the same way as in Eq. (\ref{Delta1}). 
Consequently $|A_2/A_1|$ is also restricted to the same range, 
Eq. (\ref{Range}).

We can actually find the exact solution even when we include the very 
small imaginary part of $1-\epsilon$ in $S_{11}$ due to the low-ranking 
Regge trajectories in elastic scattering. In this most general case 
it is convenient to write the $S$-matrix in the form
\begin{equation}
    S  = \left(\begin{array}{cc}
 (1-\epsilon)e^{2i\chi_1} & 
  i\sqrt{2\epsilon -\epsilon^2}e^{i(\chi_1+\chi_2)}\\
  i\sqrt{2\epsilon -\epsilon^2}e^{i(\chi_1+\chi_2)} & 
      (1-\epsilon)e^{2i\chi_2} \end{array} \right),
       \;\;\;(\epsilon^* =\epsilon). \label{SG}
\end{equation}
Unitarity and symmetry fixes the phase of $S_{12}$ as shown above
once those of $S_{11}$ and $S_{22}$ are given. For this reason one 
earlier literature\cite{Gillespie} assigned the angles $\chi_1$ 
and $\chi_2$ to the ``initial-state'' and ``final-state'' interactions 
of scattering, hinting that the phases of decay amplitudes $A_1$ and $A_2$ 
acquire $\chi_1$ and $\chi_2$, respectively, by FSI. Unfortunately this 
interpretation was wrong. As we see below, the phases of $A_{1,2}$ are 
$\chi_{1,2}$ {\em plus} additional contributions that depend on mixing 
of the channels and weak interaction. It should also be pointed 
out that the simple phase relation like $\arg S_{ij}=
\arg\sqrt{(-i)^2S_{ii}S_{jj}}$ holds only for the $S$-matrix of 
$2\times 2$, not of more than two channels.

To solve the FSI relation with Eq. (\ref{SG}) we factor out the 
``elastic phases'' $\chi_1$ and $\chi_2$ of the diagonal $S$-matrix
elements $S_{ii}$ ({\em not} of 
the partial-wave amplitudes $a_l(s)$) from the decay amplitudes by 
introducing $A'_i$ by $A_i'= e^{-i\chi_i}A_i$ ($i=1,2$). Then the 
FSI relation $A'=SA^{'*}$ reduces to the form identical to 
Eq. (\ref{S2'}). Therefore the solution for $A_{1,2}$ can be 
immediately written as Eq. (\ref{Solution2}):
\begin{eqnarray}
    A_1 &=& e^{i\chi_1}(a_1 +ib_1) \nonumber \\
    A_2 &=& \sqrt{\frac{2-\epsilon}{\epsilon}}e^{i\chi_2}\bigg(b_1
               +i\frac{\epsilon}{2-\epsilon}a_1\bigg),  \nonumber \\
  \tan(\Delta_1-\chi_1)& & \!\!\!\!\!\!\!\tan(\Delta_2-\chi_2) 
    =\frac{\epsilon}{2-\epsilon},    \label{SolutionG}
\end{eqnarray}
where $a_1$ and $b_1$ are real
The first and the second line of Eq. (\ref{SolutionG}) requires 
$|A_2/A_1|$ to remain in the same range as in the previous two 
special cases. (cf Eq. (\ref{Range})).  The relation of 
Eq. (\ref{Delta1}) is trivially modified as
\begin{equation}
   \sin^2(\Delta_1-\chi_1) =\frac{\epsilon}{4(1-\epsilon)}
   \biggl[(2-\epsilon)\frac{|A_2|^2}{|A_1|^2} - \epsilon\bigg].  
     \label{DeltaG}
\end{equation}
This relation is the bottom line of the general two-channel toy 
model: The total FSI phase $\Delta_1$ of $A_1$ is sum of the
rescattering phase $\chi_1$ of $\sqrt{S_{11}}$, not of the elastic 
partial-wave amplitude ($a_l=\frac{1}{2i}(S_l-1)$), plus the the 
phase due to rescattering through the second channel. It cannot 
be over-emphasized that in the presence of inelasticity the phase 
of $\sqrt{S_{11}}$ is 
very different from the phase of the partial-wave amplitude.  
For Pomeron-dominated scattering, for instance,  
$\arg a_0(s)=\frac{1}{2}\pi$ since $a_l(s)$ is purely imaginary for all 
$l$, but $\arg\sqrt{S_{11}} = 0$ or $\pm \pi$  
since $S_{11}$ is real and positive for $0<{\rm Im}a_0<0.5$. 
The phases of $\sqrt{S_l}$ and $a_l(s)$ would be equal only 
in the elastic limit where $S_{11}=e^{2i\delta_1}$ and 
$a_l(s)=(1/2i)(S_{11}-1) = e^{i\delta_1}\sin\delta_1$. It makes no 
sense whatsoever even as an approximation to equate the FSI phase 
with the phase of the elastic partial-wave amplitude $a_l$ in $B$ decay.
  
It is worth mentioning here that the solutions of the two-channel problem,
Eqs. (\ref{SolutionG}) and (\ref{DeltaG}), apply to the $\Omega^-$ decay
into $\Lambda K^-$ and $(\Xi\pi)_{I=1/2}$. The $\Lambda K^-$ and $\Xi\pi$
yields add up to over 99\% of the observed nonleptonic final states.  In 
this case the lower ranking Regge trajectories contribute more to the 
diagonal $S$-matrix elements than in the light hadrons channels of $B$ decay. 

While the two-channel toy model casts light on many important issues, 
it has one undesirable feature that the ratio $|A_2/A_1|$ is restricted 
within the rather narrow range set by Eq. (\ref{Range}). Numerically, 
\begin{equation}
               0.58\leq |A_2/A_1|\leq 1.73. 
\end{equation}
This constraint limits applicability of the two-channel toy-model to 
$B$ decay modes. We must extend to more than two channels to study 
$B$ decay. However, as the coupled channels 
increases, the number of dynamical unknowns quickly increases in the FSI 
relation. In order to keep our problem manageable, we must introduce 
some approximation that keeps mathematical complexity under control.

\section{Truncated multichannel problem}
   Going back to our fundamental equation Eq. (\ref{FSI}), we
consider the situation where one inelastic channel makes a dominant 
feed back to the elastic channel (channel 1) and all other inelastic 
channels are either unimportant individually or largely cancel among 
them.  We are specifically interested in the case,
\begin{equation}
   |S_{21}|\ll|S_{11}|,\;\;{\rm but}\;\; |A_2|\gg |A_1|
     \label{Assumption}
\end{equation} 
such that
\begin{equation} 
         |S_{12}A_2^*| = O(|S_{11}A_1^*|),\;\;
   |\sum_{j\geq 3}S_{1j}A_j^*| \ll |S_{12}A_2^*|. \label{Assumption2}
\end{equation}
In this case we can truncate the sum over the inelastic channels at $j=2$;
\begin{equation}
   A_1 = \sum_{j=1,2} S_{1j}A_j^* +\sum_{j\geq 3}S_{1j}A^*_j 
      \simeq S_{11}A_1^* + S_{12}A_2^*.  \label{T1}
\end{equation}
The FSI relation for the channel 2 reads 
\begin{equation}
   A_2 = S_{21}A^*_1 + S_{22}A^*_2 + \sum_{j\geq 3}S_{2j}A^*_j. 
            \label{T2}
\end{equation}
Since unlike the off-diagonal $S_{21}$ the diagonal S-matrix element 
$S_{22}=1+2ia_0^{22}(s)$ contains the term unity, we expect that
$S_{22}$ is $O(1)$ or a substantial fraction of it unless the
term $1$ is cancelled accidentally by $2ia_0^{22}(s)$ with high 
accuracy. In comparison, $S_{12}$ represents a small leakage 
into a dominant inelastic channel in the present case. Therefore 
\begin{equation}
   |S_{21}A_1^*/S_{22}A_2^*|=|S_{21}/S_{22}|\times |A_1^*/A_2^*|,
\end{equation} 
where the both factors in the right-hand side are small. In other words, 
the effect of the single elastic channel back on a robust inelastic 
channel $j$ is negligible when the coupling $S_{j1}$ between them is 
feeble. Therefore, magnitude of the first term $S_{21}A^*_1$ in the 
right-hand side of Eq. (\ref{T2}) is much 
smaller than that of $|S_{22}A_2^*|$ by the assumptions made in 
Eqs. (\ref{Assumption}) and (\ref{Assumption2}). Therefore, we may 
drop the first term in Eq. (\ref{T2}). Then no information of the 
channel 1 is needed to solve Eq. (\ref{T2}) for $A_2$. Therefore 
we solve only Eq. (\ref{T1}) and obtain a relation between $A_1$ and 
$A_2$. Solving Eq. (\ref{T2}) for $A_2$ may be hard. In the numerical
exercise later we do not attempt to compute for $A_2$ theoretically 
in terms of other inelastic channels, but resort to experiment for
information of $A_2$. 

The $S$-matrix now need not satisfy unitarity in the subsector of 
channel 1 and 2. It can be written in general as
 \begin{equation}
    S  = \left(\begin{array}{cc}
 (1-\epsilon)e^{2i\chi_1} & 
  i\kappa e^{i(\chi_1+\chi_2+\chi_{\kappa})}\\
  i\kappa e^{i(\chi_1+\chi_2+\chi_{\kappa})} & 
      \lambda e^{2i\chi_2} \end{array} \right),
      \;\; (0 < \kappa < \sqrt{2\epsilon-\epsilon^2},\;
              0< \lambda < \sqrt{1-\kappa^2} ), \label{S}
\end{equation} 
where two real parameters $\kappa$ and $\lambda$ have been introduced 
to describe inelasticity of scattering. When the channel 2 is also 
a two-body light-hadron channel, the value of $\lambda$ is 
$\simeq 1-\epsilon$ and $\chi_2\simeq \chi_1$. 
In $B$ decay the branching fractions to two-body light-mesons are much 
smaller than those to charmed meson pairs by the property of 
weak interaction. The decay $B\to K\pi$ through $D^*D^*_s$ is a typical 
example since $|\kappa|$ is much smaller than $1-\epsilon$ but $|A_2|$ 
is much larger than $|A_1|$. It has been speculated that the presence 
of the $D^*D^*_s$ channel may generate a large FSI phase for 
$K\pi$\cite{Charming}. We will examine this possibility later.

The FSI relation can be solved for $A_{1,2}$ even in the presence of 
the additional parameter $\kappa$ by the rephasing technique that we 
have used earlier. When we express $A_1$ and $A_2$ in terms of $a_2$
and $b_2$ instead of $a_1$ and $b_1$, the solution of Eq. (\ref{FSI}) 
with Eq. (\ref{S}) is:
\begin{eqnarray}
   A_1 &=& \kappa e^{i\chi_1}\bigg(\frac{1}{\epsilon}
 [-a_2\sin(\chi_2+\chi_{\kappa})+b_2\cos(\chi_2+\chi_{\kappa})]  
                          +\frac{i}{2-\epsilon}
 [a_2\cos(\chi_2+\chi_{\kappa})+b_2\sin(\chi_2+\chi_{\kappa})]\biggr) 
                       \nonumber \\
   A_2 &=& a_2 + ib_2,       \label{SolutionS} 
\end{eqnarray}
As we have pointed out, the parameter $\lambda$ is not needed 
to express the relation between $A_1$ and $A_2$ in the truncated 
approximation. Although $\chi_{\kappa}$ enters $A_1$, we can express 
the phase $\Delta_1$ without $\chi_{\kappa}$ by using experimental 
knowledge of $|A_2/A_1|$. 
Rewriting Eq. (\ref{SolutionS}) with the magnitudes and phases,
we obtain a simple generalization of the previous relation,
\begin{equation}
  \sin^2(\Delta_1-\chi_1) = \frac{1}{4(1-\epsilon)}\bigg(\kappa^2
      \frac{|A_2|^2}{|A_1|^2}-\epsilon^2\bigg). \label{SolutionT}
\end{equation}
The right-hand side of Eq. (\ref{SolutionT}) gives the contribution 
of the channel coupling that is to be added to the elastic 
contribution $\chi_1$. The ratio $|A_2/A_1|$ is now bounded as
\begin{equation}
   \frac{\epsilon}{\kappa}\leq \frac{|A_2|}{|A_1|}\leq 
         \frac{2-\epsilon}{\kappa}
                                          \label{RangeG}
\end{equation}
If $\kappa$ is small, that is, if the leakage into the channel 2
is small, large values can be accommodated for $|A_2/A_1|$. 
Therefore the truncated model is applicable to more general 
situations than the two-channel toy model that we have discussed. 
The ratio $|A_2/A_1|$ contains information of weak interaction. 
It is amusing to see in Eq. (\ref{SolutionT}) that the strong phase 
$\Delta_1$ coincides with the small ``elastic phase'' $\chi_1$ of 
$\sqrt{S_{11}}$ when $A_2$ takes the smallest value, 
$|A_2/A_1|=\epsilon/\kappa$, in the allowed range of Eq. (\ref{RangeG}).
The channel coupling effect $|\Delta_1-\chi_1|$ is the strongest
when $|A_2/A_1|$ takes the upper limit $(2-\epsilon)/\kappa$ in
Eq. (\ref{RangeG}).

If we proceed further and include two prominent inelastic channels, 
the FSI relation for $A_1$ is:
\begin{equation}
  A_1\simeq S_{11}A_1^* + S_{12}A_2^* + S_{13}A^*_3. \label{FSI3} 
\end{equation} 
If we continue along this line and incorporate more inelastic channels, 
the FSI equation for channel 1 turns into
\begin{equation}
  a_1+ib_1 = S_{11}(a_1+ib_1)^* +\sum_{j\geq 2}S_{1j}A_j^*.
                                         \label{FSIG}
\end{equation}
The first term in the right-hand side can be viewed as counteraction 
of elastic rescattering. It affects not only on the phase of $A_1$ 
but also causes long-distance enhancement or suppression on magnitude 
of $A_1$ depending on the force in the elastic channel. With 
$S_{11}=1-\epsilon$, Eq. (\ref{FSIG}) reveals an interesting 
general feature of the multichannel FSI. Separating the real and 
imaginary parts of Eq. (\ref{FSIG}), we have in the case of real
$S_{11}$
\begin{eqnarray}
 a_1 &=&\frac{1}{\epsilon}{\rm Re}\sum_{j\geq 2}S_{1j}A_j^*,
    \nonumber \\
 b_1 &=& \frac{1}{2-\epsilon}{\rm Im}\sum_{j\geq 2}S_{1j}A^*_j.
       \label{FSIab}
\end{eqnarray} 
Eq. (\ref{FSIab}) shows that elastic rescattering $S_{11}$ 
enhances the inelastic rescattering effect by $1/\epsilon$ 
$(\simeq 2)$ for the real part $a_1$ of $A_1$ and and suppresses 
it for the imaginary part $b_1$ by $1/(2-\epsilon)$ ($\simeq 
\frac{2}{3}$). This characteristic depends only on $S_{11}$, 
which we belive we know fairly accurately.

Despite its simplicity Eq. (\ref{FSIab}) contains useful information. 
For instance, {\em if} the transition to and from channel 1 can be 
described by the Born terms of $t$- and $u$-channel exchanges, 
the off-diagonal partial-wave amplitudes are real so that the
off-diagonal $S$-matrix $S_{1j}$ $(j \geq 2)=2ia_0^{(1j)}$ are purely 
imaginary. Therefore the real decay amplitudes $A_j$ ($j\geq 2$) of 
the inelastic channels contribute to the imaginary part $b_1$ of 
channel 1 in this case. Even if there is no resonance in the process, 
the phase $\Delta_1$ can be very large in this way. If furthermore 
the ``inelastic'' decay amplitudes $A_j$ ($j\geq 2$) happen to be 
all real, the phase $\Delta_1$ would be $\pm 90^{\circ}$. This is not 
surprising: In the language of dispersion theory the on-shell inelastic 
intermediate states generate an absorptive part for $A_1$ that 
turns out to be purely imaginary in this situation.  Our Eq. 
(\ref{FSIab}) involves one dynamical input: The elastic scattering 
amplitudes of light mesons are almost purely imaginary 
(Pomeron-dominated). 

\section{Numerical exercise}

   Some numerical exercise is called for to show relevance of our 
endeavor to the $B$ decay of the real world. Contrary to the initial 
optimism that had prevailed before $B$ physics experiment started, 
analysis of experiment seems to suggest that the FSI 
phases of some two-body decay amplitudes appear to be much larger than 
what we expected in the original short-distance picture\cite{Neubert}. 
The word $K\pi$ puzzle has been coined for the unexpectedly large 
tree-contribution and/or FSI phases in $K\pi$ modes. As the perturbative
technique has become more sophisticated, people have come to agree 
that emission and absorption of soft and collinear quarks and gluons 
plays an important role in many decay modes\cite{Softcol,Chay}. Such soft 
constituents contribute to the FSI phases involving long-distance physics. 
While one can parametrizes such contributions in the soft-collinear 
theory, one cannot evaluate them numerically in perturbative argument. 
Our $S$-matrix approach also has its own drawback: While elastic 
scattering of two light hadrons at energy $m_B$ has been well understood, 
we know less about their inelastic scattering. Nonetheless we would like 
to show here that our method may be useful in some of $B$ decay modes. 

It is believed that the decay $B\to K\pi$ occurs primarily with 
the penguin interaction $\sim [(\overline{b}s)(\overline{q}q)+h.c]$. 
In the penguin process the coupling to channels such as $K^*\rho$ 
need to be studied as a source of the strong phase of the $K\pi$ 
amplitude.  However, it has been argued that $K\pi$ can be 
produced indirectly with the tree interaction 
$\sim [(\overline{b}c)(\overline{c}s)+ {\rm h.c.}]$  as well through 
the charmed-meson-pair states such as $DD_s$ and $D^*D^*_s$\cite{Charming}. 
The $K\pi$ amplitude of this process acquires a strong phase different from 
the direct penguin amplitude.  What can our approach say about this problem ?

The branching fractions have been measured for the following
two-body channels that couple to $K\pi$ in $B^0$ 
decay\cite{PDGweb}:
\begin{eqnarray}
     B(K^+\pi^-) &=& (1.88\pm 0.07)\times 10^{-5}, \nonumber \\
     B(K^0\pi^0) &=& (1.15\pm 0.10)\times 10^{-5}, \nonumber \\
     B(K\eta)&<& 2.0 \times 10^{-6},\\
     B(K\eta')&=& (6.5\pm 0.4)\times 10^{-5},\\
     B(K^{*0}\phi) &=& (0.95 \pm 0.08)\times 10^{-5}\nonumber\\
     B(K^{*+}\rho^-) &<& 1.20 \times 10^{-5}\nonumber \\
     B(K^{*0}\omega) &<& 4.2 \times 10^{-6}\nonumber \\
     B(D^-D_s^+) &=& (6.5\pm 1.3)\times 10^{-3},\nonumber\\
     B(D^{*-}D_s^{*+}) &=& (1.77 \pm 0.14)\times 10^{-2}. 
                        \label{BR}
\end{eqnarray}
All of them can make $J^{P}=0^+$, the spin-parity of $K\pi$.  
The states $K\pi$ and $K^*\rho$ can be either in $I=\frac{1}{2}$ or 
$I=\frac{3}{2}$ while all other modes are only in $I=\frac{1}{2}$.

From the Regge phenomenology on elastic $K\pi$ scattering,
we have already estimated $\epsilon$ for the Pomeron contribution
in Section III. The $\rho$ and $f_2$ trajectories with exchange 
degeneracy allow us to estimate the small imaginary part of $S_{11}$.
By adding the $\rho$ and $f_2$ Regge contributions, we obtain 
numerically
\begin{equation}
    S_{11} \simeq \left\{ \begin{array}{cc}
         0.39\times e^{0.06i}& (I=\frac{1}{2}),\\
         0.46\times e^{-0.10i}& (I=\frac{3}{2}). \label{Nonleading}
         \end{array} \right.
\end{equation}
The $\rho/f_2$ contributions to the scattering amplitudes $T$ are dominantly 
imaginary for $I=\frac{1}{2}$ and real for $I=\frac{3}{2}$, as we expect from
$s$ to $t$-channel duality. Therefore
they generate a larger imaginary part for $S=1+2iT$ in the $I=\frac{3}{2}$
channel. This is very different from our intuitive picture in the 
single-channel case. Though we do not attach errors to our estimate, 
errors as large as factor two are possible for the phases in Eq. 
(\ref{Nonleading}).
 
Estimate of $S_{12}$ is less reliable than $S_{11}$ owing to very
indirect experimental information and to larger theoretical 
uncertainties. We eliminate the $K^*\phi$ channel from our consideration 
since the Regge residue or the coupling of $\phi$ to $\pi\rho$ 
is highly suppressed (``OZI-forbidden''). In contrast $K^*\rho$ can 
couple to $K\pi$ without such suppression. While a loose upper bound 
has been set on $K^*\rho$ experimentally (see Eq. (\ref{BR})), we believe 
that the $K^*\rho$ channel is more important than the $K^*\phi$ channel. 
We may use the Regge phenomenology to estimate $S_{12}$ for the $K^*\rho$
state of longitudinal polarizations in the final state. For the
scattering $K\pi \to K^*\rho$, the leading Regge poles are $\omega$ and 
$a_2$ which are exchange degenerate.\footnote{
The $K^*$ trajectory generates a backward peak is generated, but it
is less important than the forward peak. We neglect the backward peak 
contribution to $S_{12}$.} 
The couplings of $\pi\rho\omega$ and $\pi\rho a_2$ are known on the mass 
shells of $\omega$ and $a_2$, respectively, from low-energy spectroscopy. 
The corresponding $KK^*$ couplings of $\omega$ and $a_2$ are
obtained by an SU(3) rotation. But we need to extrapolate them to
the off-shell $\rho$ and $a_2$ to relate them to the Regge 
residues. This extrapolation is a major source of uncertainty. If 
we ignore the extrapolation, they are at the same level in magnitude
as the nonleading Regge contributions in $S_{11}$; 
\begin{equation}
  S_{12}^{K^*\rho} \approx \left\{ \begin{array}{cc}
         -0.07 + 0.02i & (I=\frac{1}{2}),\\
         -0.05i  & (I=\frac{3}{2}).
         \end{array} \right. \label{rhoKstar}
\end{equation}
Using the ratio $|A_2/A_1|$ computed with the measured branching 
fraction and the upper bound listed in Eq. (\ref{BR}), we reach
the crude estimate,
\begin{equation}
   |S_{12}^{K^*\rho}A_2^*| < 0.1 \times |S_{11}A_1^*|,
            \label{Estimate}
\end{equation}
Despite large uncertainty of these numbers  we may conclude
with Eq. (\ref{rhoKstar}) that the inelastic term 
$S_{12}A_2^*$ of the $K^*\rho$ channel in the $K\pi$ mode is not
significant relative to the elastic term $S_{11}A_1^*$. It is 
certainly not a major source of the strong phase for the $K\pi$
amplitude in the standard penguin decay (not through $c\overline{c}$). 
The same line 
of estimate suggests that the $K^*\omega$ state is neither important to 
the FSI phase of $K\pi$. Some may wonder about one-pion exchange in $K\pi
\leftrightarrow K^*\rho$. The Reggeized pion exchange amplitude for 
$K\pi\to K^*\rho$ is down by another power of $s^{0.5}$ relative to 
those of $\omega$ and $a_2$ exchanges. The small denominator of the 
pion propagator does not enhance the amplitude near forward direction 
because the Lorentz structure requires the amplitude near the pion 
pole to be proportional to
\begin{equation}
 \frac{(\epsilon^{(\rho)}\cdot p_{\pi})(\epsilon^{(K^*)}\cdot p_K)}{
              m_{\rho}^2-2(p_{\pi}\cdot p_{\rho})}.
\end{equation} 
The both factors in the numerator vanish up to $O(4m_{\rho}^2/m_B^2,
4m_{K^*}^2/m_B^2)$ for longitudinally polarized $K^*$ and $\rho$  
in the forward direction (${\bf p}_{\rho}\parallel{\bf p}_{\pi}$). 
They eliminate a forward peak from the pion pole. The Regge theory 
predicts that the rest of the scattering amplitude falls sharply 
($ \sim e^{\alpha'_{\pi}(0)t\ln s}$) off the forward direction. 
Therefore the pion exchange can be dismissed. Then we feel safe to 
conclude with Eq. (\ref{Estimate}) that 
the coupling of $K\pi$ to the $K^*\rho$ channel is not important in 
determining the FSI phases for the $K\pi$ modes. 
 
The $K\eta$ and $K\eta'$ channels can couple to the $K\pi$ channel of
$I=\frac{1}{2}$ through the $a_2$ Regge exchange. Although this 
contribution has the same $m_B$ dependence as the $\rho$ and $f_2$
Regge exchanges, the $a_2$ Regge residues with $\pi\eta(\eta')$ are
most likely smaller than the residue with $\pi\rho$: We can 
estimate from the $a_2$ decay branching that, after the $d$-wave 
phase space factor is separated, the on-shell $a_2$ couplings to 
$\pi\eta$ and $\pi\eta'$ are about factor 20 smaller than those 
to $\pi\rho$. Therefore neither $K\eta$ nor $K\eta'$ compete with
$K^*\rho$ in the final state. Therefore we may leave out $K\eta$
and $K\eta'$ from our consideration.

The contributions from nonresonant three-body final states are harder 
to estimate since computation of $S_{12}$ is next to impossible. 
There are many nonresonant multiparticle channels with relatively 
minor branching fractions. If rescattering of $K\pi$ to multibody channels
is quasi-diffractive with no quantum number exchange, the Pomeron can
contribute. In such scattering the final states consist of two lumps
of relatively small invariant masses that carry the same flavors
as $K$ and $\pi$. They are likely to end up in two-body states of highly 
excited meson states, for instance, $K_2(1430)a_1$. While this is a 
possibility, none of such modes have been positively identified so far 
in measurement.

Genuine nonresonant three-body channels are probably not a major 
source of the FSI phases, unless their contributions add up by 
constructive interference to a large value. In fact, it is 
conceivable that they sum up in random 
phases into relatively a small number\cite{CS,SW}. That is one 
motivation when we have introduced the truncated approximation.
Our tentative conclusion on the $K\pi$ amplitudes ($I=\frac{1}{2}, 
\frac{3}{2}$) of the light-quark penguin decay operators is that 
the FSI phase produced by coupling to the inelastic channels is 
insignificant.  Analysis of the $K\pi$ amplitudes in search of the
weak phases was started more than ten years ago. With little
knowledge of the strong phases, however, the analysis could be
carried out only by assuming that the long-distance strong phases 
be negligible\cite{Falk}. 

   The charmed meson-pair channels are very different. Since they are 
the CKM dominant tree-decay final states, their branching fractions 
are a few orders of magnitude larger than that of the penguin-dominated 
decay. They can annihilate into $K\pi$. In the quark picture this 
process can be viewed as the on-shell contribution of the $c\overline{c}$ 
penguin to $K\pi$. Some call this process as 
``charming penguin''\cite{Charming}. Among the charmed meson pairs, 
the most prominent decay channel of $J^P=0^+$ is $D^*D_s^*$. 
Its amplitude can be estimated with Eq. (\ref{BR}) as
\begin{equation}
     |A_{D^{*-}D_s^{*+}}/A_{K\pi(I=1/2)}| \simeq 25. \label{BrBr}
\end{equation}
One important question here is how much of the observed total 
$A_{K\pi(I=1/2)}$ is the ``charming penguin'' contribution. Since
we expect that long-distance physics enters the on-shell process of 
charmed-meson pairs at energy $m_B$, it is not easy to evaluate its 
magnitude. Some argue that it can be very large\cite{Bauer,Bauer2}. 
However, a counter argument was made to advocate the short-distance
argument\cite{Beneke2}. Theorists have not come to consensus 
on magnitude of this contribution. Therefore we insert one fudge factor 
$r$ here for this contribution to $A_1$ as
\begin{equation}
     |A_{D^{*-}D_s^{*+}}/A_{K\pi(I=1/2 \; via\; c\overline{c})}| 
                             \simeq 25\times\frac{1}{r}, \;\; (r<1)
\end{equation}
where $r$, the fraction of the $c\overline{c}$ contribution, may be 
as large as a half or even more\cite{Bauer}. It must be settled by theory 
rather than by experiment. With this fudge factor Eq. (\ref{SolutionT}) 
turns into
\begin{equation}
  \sin^2(\Delta_1-\chi_1) = \frac{1}{4(1-\epsilon)}\bigg(\kappa'^2
      \frac{|A_2|^2}{|A_1|^2}-\epsilon^2\bigg), \label{SolutionT'}
\end{equation}
where $\kappa' = \kappa/r=|S_{12}/rS_{11}|$.
Even if the spill-over of $D^*D_s^*$ into $K\pi$ is as tiny as
one tenth of percent ($\kappa^2 = |S_{12}/S_{11}|^2\simeq 10^{-3}$), 
the $D^*D_s^*$ channel may control the FSI phase of the $K\pi$ 
amplitude that comes through $c\overline{c}$. To proceed further, 
we must look into the transition $K\pi\leftrightarrow D^*D_s^*$. 
(Fig.4)
\begin{figure}
\epsfig{file=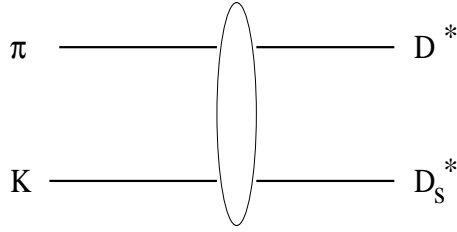,width=6cm,height=3cm} 
\caption{The dominant inelastic scattering $K\pi\to D^*D^*_s$.  
\label{fig:4}}
\end{figure}

  Application of the Regge theory is questionable to the charm-pair
channels since the charmed meson masses are around 2 GeV 
and the total energy is a little above 5 GeV. Departing from 
the Regge theory, let us make the Born approximation in $t$-channel 
exchange. The exchanged mesons are the charmed mesons $D$, $D^*, 
\cdots$. For $D$-exchange, we know the $D^*\overline{D}\pi$ coupling from 
the decay $D^*\to D\pi$ and can compute the $D^*D_sK$ coupling 
by an SU(3) rotation of the $D^*\overline{D}\pi$ coupling. The 
differential cross section rises toward the forward direction
when the $D$ meson is exchanged.  For $D^*$ exchange, we can deduce the 
$D^*\overline{D}^*\pi$ coupling from the $D^*\overline{D}\pi$ coupling 
with the heavy quark spin symmetry and rotate it into the $D^*D_s^*K$ 
coupling by SU(3). In contrast to the $D$ exchange, the Lorentz structure 
of the vertex $\sim\varepsilon^{\mu\nu\kappa\lambda}\epsilon_{\nu}
p_{1\kappa}p_{2\lambda}$ of $D^*$ exchange cancels a forward peak 
that would be otherwise generated by the $D^*$ propagator. For this 
reason the $D^*$-exchange is less important. When we compute the $D$ 
contribution in the Feynman diagram with the on-shell couplings, we 
obtain $|S_{12}|\approx 0.5$.  But this is obviously a nonsense. The reason 
is that we have ignored the form-factor damping effect of the exchanged 
off-shell $D$. For an order-of-magnitude estimate we may multiply a factor 
of $m^2_*/m_{D}^2$ as a form-factor effect where $m_*\simeq 0.3\;{\rm GeV}$ 
is the binding scale of the charmed mesons. Then our estimate goes 
down by more than an order of magnitude from $|S_{12}|\approx 0.5$ to 
$|S_{12}|\approx 0.014$. This 
latter value is probably closer to reality. It is roughly in line with 
the rule of thumb; in the quark picture a pair creation probability 
of $c\overline{c}$ is suppressed by about $(m_q/m_c)^2$ relative to 
light-quark pair creation of $q\overline{q}$, where the quark masses 
are the constituent masses. This rule works roughly for $s\overline{s}$ 
and $c\overline{c}$ production in high-energy collision. If we use this 
rule, we obtain $|S_{12}|\simeq 0.01$ from Eq. (\ref{rhoKstar}) with 
$m_s/m_c\simeq 1/3$. Therefore we choose as our best guess
\begin{equation}
    S_{12}^{D^*D_s^*} \approx 0.01i. \label{DDs} 
\end{equation}
As we have noted earlier, $S_{12}$ is purely imaginary for real $a_0(s)$. 
The number of Eq. (\ref{DDs}) is obviously an order-of-magnitude estimate
at best. With $\epsilon\simeq 0.5$ and tentatively $r\approx 0.5$, 
a crude central value of our estimate for $\kappa'$ is
\begin{equation}
    \kappa'\approx 0.01/[0.5\times(1-0.5)]= 0.04. \label{DDs2}
\end{equation}
We now substitute all these numbers in Eq. (\ref{SolutionT'}) of 
the truncated approximation.  We take the number of Eq. (\ref{DDs2}) 
as a ballpark figure and sweep the value of $\kappa'^2$ by a factor 
two across this value to see 
what FSI phase can be generated for $K\pi$ of $I=\frac{1}{2}$ by the 
channel coupling to the $D^*D_s^*$ channel. The result is plotted in 
Fig. 5. The value of $\kappa'|A_2/A_1|$ is constrained between 0.5 and 1.5
by Eq. (\ref{SolutionT'}) and sweeps in the region between two vertical 
broken lines. 

\begin{figure}
\epsfig{file=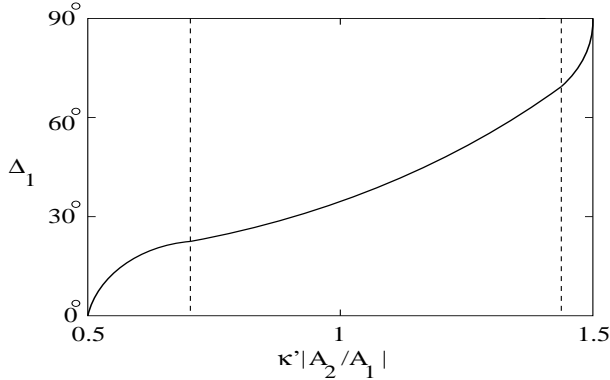,width=8cm,height=5cm} 
\caption{The FSI (strong) phase of the $K\pi$ decay channel of
$I=\frac{1}{2}$ as the $K\pi\to D^*D_s^*$ transition amplitude is 
varied in magnitude.  
\label{fig:5}}
\end{figure}
 
Within the uncertainty of $\kappa'$, the FSI phase $\Delta_1$ of 
the $K\pi$ amplitude through $c\overline{c}$ can be any value between 
$21^{\circ}$ and $69^{\circ}$. Eq. (\ref{SolutionT'}) does not 
determine the sign of $\Delta_1$ since it does not contain full 
information of $A_2$. Because of the large uncertainty of $\kappa'$,
we cannot constrain $\Delta_1$ meaningfully at present.  Even $\Delta_1 
\simeq 90^{\circ}$ is not reliably excluded. Keeping the uncertainty of our 
estimate in mind, we should state here only that the channel coupling to 
$D^*D^*_s$ is capable of producing a very large strong phase for the 
charming penguin $K\pi$ amplitude, in particular in the case that 
the ``charming penguin amplitude'' is a sizable fraction of the total 
amplitude. Quantitatively reliable computation of the FSI phase will 
be possible after we have obtained a better theoretical estimate 
of $S_{12}$ for $K\pi\leftrightarrow D^*D_s^*$ as well as magnitude 
of the decay amplitude through $c\overline{c}$. 
    Until then we must not set this strong phase to zero but
leave it as an unknown parameter to be determined by fit to 
experimental data. If such experimental fit clearly requires a large 
FSI angle for the $K\pi$ mode of $I=\frac{1}{2}$ but not of $I=\frac{3}{2}$, 
we shall be able to assert that the on-shell $c\overline{c}$ intermediate 
state plays an important role in the decay $B\to K\pi$.   

The statement above can be made for many other two-body light-hadron 
modes. The $\pi\pi$ mode couples to $\rho\rho$ whose branching fraction is
nearly an order of magnitude larger than that of $\pi\pi$. But perturbative 
calculation of the $\pi\pi\leftrightarrow\rho\rho$ transition\cite{Vysotsky} 
by the Born diagrams without off-shell damping can easily overestimate it. 
With an estimate of $S_{12}$ along the same line as for $K\pi\leftrightarrow 
K^*\rho$, the right-hand side of Eq. (\ref{SolutionT}) comes out to be 
negative in the case of $\pi\pi\to \rho\rho$, It 
implies that the branching fraction of $\rho\rho$ is not large enough 
to affect the FSI phase for the $\pi\pi$ amplitudes of the tree decay 
$\sim (\overline{b}u)(\overline{q}{q})+{\rm h.c.}$ ($q$ = light quarks).  
The $\pi\pi$ of $I=0$ can be fed also by the charmed hadron channel 
$D^*\overline{D}^*$ whose branching fraction is two orders of 
magnitude larger than that of $\pi\pi$. For the $\pi\pi$ amplitude
through $\overline{c}c$, coupling to $D^*\overline{D}^*$ is the most 
important source of the strong phase. To obtain the strong phases and
to compare with experiment, we again need to know about relative importance 
of the two classes of decay; $b\to u\overline{u}d$ 
and $b\to c\overline{c}d$ for $B\to\pi\pi$.
 
In contrast, the tree-dominated CKM-favored decay modes such as 
$\overline{B}\to D\pi$ have no wide open inelastic channels. Transition 
to the $D^*\rho$ channel is as insignificant as the transition to
$K\pi\to K^*\rho$ in the $K\pi$ mode. Since the $D^*\rho$ decay branching 
fraction is comparable to that of $D\pi$ and $S_{12}$ is much smaller 
than $S_{11}$ (cf Eqs. (\ref{rhoKstar}) and (\ref{Estimate})), of its 
elastic scattering, we expect that $|S_{12}A_2^*|\ll |S_{11}A^*_1|$ 
and therefore the channel coupling contribution is unimportant. 
It means that the FSI phases are small in these modes and that 
simple short-distance calculation of the phases can produce 
an answer not far from reality. 

\section{Comment and discussion}
  One of our purposes is to solve the most general two-channel model
exactly and to clarify the mechanism of generating strong phases in this 
toy model. We have seen above that existence of competing channels 
completely changes the strong phase from that of the ``elastic 
rescattering phase'' in all cases of two-channel $S$-matrix. 
Unitarity plays an important role here.  The other 
purpose is to introduce a feasible approximation scheme which may be 
applicable to cases of special interest in $B$ decay.  We have 
applied this method to the decay $B\to K\pi$ and have made semiquantitative 
analysis. But its outcome is not numerically satisfying because of 
limitation in available knowledge about off-diagonal scattering.
We have extended our analysis to the general multichannel case and
come to one simple interesting observation: Though it may sound odd, 
inelastic scattering tend to enhance the strong phase of the elastic
channel most when inelastic amplitudes are real ({\em i.e.,} $S_{1j}
=iT_{1j}$ = imaginary) 
rather than imaginary, if the strong phases of the inelastic decay 
amplitudes are small. (cf Eq. (\ref{FSIab}).)

  It is a big challenge to go beyond the two-channel approximation.
As the number of channels $n$ increases, the number of parameters
in the FSI relation increases as $\frac{1}{2}n(n+1)$ for strong
interaction. In addition we need piece of information from weak 
interaction. In the truncated approximation we have kept only 
a single dominant one among the inelastic channels. Our assumption 
is that all other inelastic channels are less important or else sum 
up in random phases to become numerically insignificant. If the 
higher inelastic channels do not sum randomly, there must be some 
underlying dynamical reason for it. In such a case some approach 
orthogonal to ours may have advantage. For instance, it is 
an approach based on quarks and gluons instead of hadrons. 

The major source of the strong (FSI) phase in the
multichannel case is the transition to inelastic channels. 
Consequently accurate computation of the FSI phases depends on 
knowledge of the transition $S$-matrix at energy $m_B$ between 
a channel of our interest and dominant inelastic channels. That is, 
we need to know dynamics of long and intermediate distances at this 
energy. The present author is of the opinion that quantitatively we 
have a little better handle on hadron physics numerically than on the 
soft and collinear quarks and gluons in this territory. But opinions 
probably divide among physicists of different generations.
 
\acknowledgments
The author is grateful to C. W. Bauer for updating him in the 
theoretical status of the charming penguin.  
This work was supported by the Director, Office of Science, Office of
High Energy and Nuclear Physics, Division of High Energy Physics,
of the U.S.  Department of Energy under contract DE--AC02--05CH11231. 

\appendix

\section{Real orthogonality of transformation}
   Expansion of the observable state $|j^{\rm out}\rangle$ in the $S$-matrix
eigenstates $|a^{\rm out}\rangle$ is defined by
\begin{equation}
      |j^{\rm out}\rangle = \sum_a O_{ja}|a^{\rm out}\rangle .
               \label{Expansion1}
\end{equation}      
At this stage the matrix ${\cal O}$ is assumed to be only unitary, not 
necessarily orthogonal. Make time-reversal on the scattering amplitude 
from eigenstate $|a\rangle$ to observable state $|j\rangle$ in our phase 
convention of states under time reversal:
\begin{equation}
      \langle j^{\rm out}|a^{\rm in}\rangle = 
      \langle j|S|a\rangle \stackrel{T}{=}
      \langle a|S|j\rangle = \langle a^{\rm out}|j^{\rm in}\rangle.
                              \label{hybridT} 
\end{equation}
Operating $S^{\dagger}$ on Eq. (\ref{Expansion1}), we obtain
\begin{equation}
     |j^{\rm in}\rangle = \sum_a O_{ja}|a^{\rm in}\rangle.
       \label{Expansion2} 
\end{equation} 
Substitution of Eqs. (\ref{Expansion1}) and (\ref{Expansion2}) in Eq. 
(\ref{hybridT}) gives us
\begin{equation}
      O^*_{ja}e^{2i\delta_a} = O_{ja} e^{2i\delta_a},
\end{equation}
which proves that ${\cal O}$ is orthogonal:
\begin{equation}
          O^*_{ja} = O_{ja}.
\end{equation}

\end{document}